\documentclass[12pt]{article}
\usepackage{amsmath,epsfig,psfrag}

\begin{document}

\begin{titlepage}

\begin{flushright}
KA-TP-1-2008\\
MZ-TH/08-1\\[0.1cm]
January 2, 2008
\end{flushright}

\vspace{0.8cm}
\begin{center}
\Large\bf
A model for the very early Universe
\end{center}

\vspace{0.5cm}
\begin{center}
{\sc Francesco Giacosa$^a$, Ralf Hofmann$^b$, and Matthias Neubert$^c$}\footnote{On leave from Laboratory for Elementary-Particle Physics, Cornell University, Ithaca, NY 14853, U.S.A.}\\[0.4cm]
{\sl $^a$\,Institut f\"ur Theoretische Physik, 
Universit\"at Frankfurt\\ 
Max von Laue-Str.\ 1, D-60438 Frankfurt am Main, Germany\\[0.3cm]
$^b$\,Institut f\"ur Theoretische Physik, 
Universit\"at Karlsruhe (TH)\\ 
Kaiserstr.~12, D-76128 Karlsruhe, Germany\\[0.3cm]
$^c$\,Institut f\"ur Physik (THEP), 
Johannes Gutenberg-Universit\"at Mainz\\ 
Staudingerweg 7, D-55128 Mainz, Germany}
\end{center}

\vspace{1.0cm}
\begin{abstract}
\vspace{0.2cm}
\noindent 
A model with $N$ species of massless fermions interacting via (microscopic) gravitational torsion in de Sitter spacetime is investigated in the limit $N\to\infty$. The $U_V(N)\times U_A(N)$ flavor symmetry is broken dynamically irrespective of the (positive) value of the induced four-fermion coupling. This model is equivalent to a theory with free but massive fermions fluctuating about the chiral condensate. When the fermions are integrated out in a way demonstrated long ago by Candelas and Raine, the associated gap equation together with the Friedmann equation predict that the Hubble parameter vanishes. Introducing a matter sector (subject to a finite gauge symmetry) as a source for subsequent cosmology, the neutral Goldstone field acquires mass by the chiral anomaly, resulting in a Planck-scale axion. 
\end{abstract}
\vfil

\end{titlepage}

\section{Introduction}

The cosmological constant problem can be divided into three questions: A) Why did no relic vacuum energy density of Planckian magnitude survive the Planckian epoch until today? B) What is the mechanism that prevents the regeneration of vacuum energy in terms of the quantum fluctuations inherent to the matter sector of the Universe for sub-Planckian cosmology? C) Why is today's value of the vacuum energy density nonvanishing yet extremely small compared to particle physics scales? If cosmological inflation indeed has occurred with a mean vacuum energy density of Planckian magnitude, then problem~A addresses the question what the mechanism for the complete extinction of inflation is. It is suggestive that consequences of a subtle symmetry principle are at work here. Problem~B may pose itself due to our insufficient understanding of the nature of elementary particles and their ground state (for a related discussion see \cite{Hofmann2005,Hofmann2007}). If a symmetry is responsible for the solution to problem~A, then problem~C must be related to a slight and explicit violation of this symmetry. Indeed, as discussed in \cite{GH2005}, the solution to problem~C may be rooted in the chiral anomaly \cite{Adler,JackiwBell,Wilczek,Peccei}, which by virtue of topologically nontrivial field configurations of Yang-Mills theories breaks an exact, nonlinearly represented global symmetry $U_A(1)$. The existence of a Planck-scale axion field (with Peccei-Quinn scale $\sim M_P$, see also \cite{frieman1995}) in connection with an $SU(2)$ Yang-Mills theory describing photon propagation may be responsible for the thus far unexplained components entering the equation of state of the present Universe \cite{GH2005}. While the existence of a flavor-neutral chiral pseudo-Goldstone field had to be assumed in \cite{GH2005}, we will show in the present work how this field naturally 
emerges as a consequence of gravitationally induced, chirally-invariant fermion interactions. At the same time, the consequences of these interactions address problem~A. Interesting ideas in this respect, which in fact have stimulated the present work, are expressed in \cite{Prokopec2006} (for a review see \cite{Nobbenhuis} and references therein). Ideas on how gravitationally induced chiral symmetry breaking could possibly relax a pre-exisitng cosmological constant can also be found in \cite{Alexander:2005vb,Alexander:2006we}.

In this article we study the cosmological implications of gravitationally induced, chirally-invariant four-fermion interactions in four dimensions. In \cite{Perez2006} it was shown that a quartic interaction of the form 
\begin{equation}\label{induced}
   -\frac{3\pi G}{2}\,\frac{\gamma^2}{\gamma^2+1} 
   \left( \bar\psi\gamma_5\gamma_a\psi \right)^2 ,
\end{equation}
where $G$ is Newton's constant and $\gamma$ the Immirzi parameter, arises when the Holst action \cite{Holst1996} is used to classically eliminate the torsion-induced, nonlocal interaction between fermions. Here we consider a scenario in which $N$ massless (chiral) fermions, $\psi^T=(\psi_1,\psi_2,\dots,\psi_N)$, with a flavor symmetry $U_V(N)\times U_A(N)$ are present in the very early Universe. We work in a de Sitter spacetime sourced both by the fermion dynamics and a
bare cosmological constant. We first address the question whether gravity can generate a chiral condensate of these fermions. By applying a Fierz transformation to the fermion fields, we show that a chiral condensate emerges if $\gamma $ is purely imaginary and $|\gamma|<1$, in which case the interaction (\ref{induced}) is attractive. In fact, when considering the cosmological evolution in the sub-Planckian regime ($H<M_{P}$, with $H$ the Hubble parameter and $M_P=(\frac{3}{8\pi G})^{1/2}$ the reduced Planck mass), only the attractive scalar channel matters. Apart from flavor-nonsinglet fields, a massive scalar isosinglet field and a massless pseudoscalar
isosinglet (Goldstone) field emerge. The massive scalar plays the dominant role in the cosmological evolution. As a consequence, we observe that for $H<M_P$ any positive vacuum energy density vanishes in the large-$N$ limit. The mean-field approximation underlying our analysis of the gap equation becomes exact in this limit. 

In Section~\ref{FT} we briefly recall how a gravitationally induced, chirally invariant, and local four-fermion interaction emerges starting with the Holst action. We then apply a Fierz transformation to this interaction and study the properties of the scalar and pseudoscalar channels. Subsequently we integrate out the fermions in a de Sitter background and consider the minimum of the emerging effective potential. We also discuss possible regularization schemes and how they influence the properties of the effective potential. In Section~\ref{cosmo} we (algebraically) solve the Friedmann equation for de Sitter spacetime and show that in the large-$N$ limit the Planckian vacuum energy density is driven to zero. We then discuss the possible role of the flavor-singlet Planck-scale axion field in late-time cosmology. Finally, in Section~\ref{SC} we summarize our work and present some conclusions.

\section{Theoretical set-up}
\label{FT}

\subsection{Gravitationally induced four-fermion interactions}
\label{subsec:4fermi}

We consider purely gravitational dynamics as given by the Holst action \cite{Holst1996}
\begin{equation}\label{faction}
   S[e,A] = \frac{1}{16\pi G} \left( \int d^4x\,
   e\,e_a^\mu\,e_b^\nu\,F_{\mu\nu}^{ab} 
   - \frac{1}{\gamma} \int d^4x\,e\,e_a^\mu\,e_b^\nu\, 
   \tilde{F}_{\mu\nu}^{ab} \right) ,
\end{equation}
which is a functional of the tetrad field $e_\mu^a$ \cite{Ashtekar1991}. Here $a=0,1,2,3$ is the internal Lorentz index, $\mu=0,1,2,3$ the coordinate index, and $e\equiv\det e^\mu_a$. $F_{\mu\nu}^{ab}$ is the curvature of the connection $A_{cd}^\mu$ defined as 
\begin{equation}\label{conn}
   A_{cd}^\mu 
   = e_c^\nu \left( e_{d,\nu}^\mu - \Gamma_{\rho\nu }^\mu\,e_d^\rho
   \right) ,
\end{equation}
and $\Gamma_{\rho\nu}^\mu$ are the Christoffel symbols. $\tilde{F}_{\mu\nu}^{ab}=\frac12\epsilon^{ab}_{\ \ cd}\,F_{\mu\nu}^{cd}$ is the (internal) dual field strength. Note that the Immirzi parameter $\gamma$ can either be real (leading to the Barbero connection \cite{Barbero1995}) or imaginary ($\gamma=i$ leads to the self-dual Ashtekar connection \cite{Ashtekar1991}).

The first term in (\ref{faction}) yields the tetrad formulation
(Palatini action) of the Einstein-Hilbert action, the latter emerging when inserting a solution to the associated equation of motion ($A_{cd}^\mu$ being a torsion-free spin connection $\omega_{cd}^\mu[e]$) into (\ref{faction}) and using $g_{\mu\nu}=e_\mu^a\,e_{\nu a}$. The second term is identically zero due to the Bianchi identity for the Riemann tensor. It follows that, regardless of the value of $\gamma$, the action (\ref{faction}) is \emph{classically\/} equivalent to the familiar Einstein-Hilbert action \cite{Perez2006}. This is reminiscent of the $\theta$-angle in Quantum Chromodynamics (QCD), which parameterizes the contributions of topology-changing fluctuations to the partition function. In QCD the associated part of the action does not enter the equations of motion if the absence of boundary terms can be assumed. As is the case with $\theta$ in QCD, different choices of $\gamma$ in (\ref{faction}) are physically not equivalent at the quantum level \cite{Rovelli1997}.

What holds true for pure gravity is no longer valid if minimally coupled chiral fermions are introduced. The equation of motion for the
tetrad $e_a^\mu$ subject to a fermionic source is solved in terms of a
connection $A_{cd}^\mu$ having two contributions, a torsion-free spin
connection for $e_a^\mu$ (as in the purely gravitational case) and a
torsion term related to the axial fermion current. Upon substituting $A_{cd}^\mu$ back into the action, a four-fermion interaction of the following form emerges \cite{Perez2006}: 
\begin{equation}\label{fermint}
   S_{\mathrm{int}} = \frac{K}{2} \int d^4x\,e \left(
    \bar\psi\gamma_5\gamma_a\psi \right) 
    \left( \bar\psi\gamma_5\gamma^a\psi \right) \,; \qquad 
   K = - 3\pi G\,\frac{\gamma^2}{\gamma^2+1} 
    = - \frac{9}{8M_P^2}\,\frac{\gamma^2}{\gamma^2+1} \,.
\end{equation}
Thus the Immirzi parameter acquires physical relevance through the presence of massless fermions, even though gravity is still treated classically. Notice that $K$ becomes positive for imaginary $\gamma$ with $|\gamma|<1$, and that it diverges for $\gamma\to\pm i$.

\subsection{Effective action after integrating out \boldmath$\psi$\unboldmath}

The action describing the fermions reads 
\begin{equation}
   S_{\mathrm{ferm}} 
   = \int d^4x\,e \left[ \bar\psi\,ie_a^\mu\gamma^a D_\mu[e]\,\psi 
   + \frac{K}{2} \left( \bar\psi\gamma_5\gamma_a\psi \right) 
   \left( \bar\psi\gamma_5\gamma^a\psi \right) \right] ,
\end{equation}
where $D_\mu[e]$ is the covariant derivative with respect to the connection $A$. We study the system of interacting fermions in a de Sitter spacetime in FRW coordinates 
\begin{equation}  \label{metric}
ds^2 = dt^2 - a^2(t)\,d\vec{x}\cdot d\vec{x} \,,
\end{equation}
where $a=a_{0}\,e^{Ht}$ is the scale factor. In this case the vierbein reads 
\begin{equation}\label{tetrad}
   e_{\mu a} = \delta_{\mu 0}\,\delta_{a0} - a(t)\,\delta_{\mu i}\,
   \delta_{a i} \,.
\end{equation}
The consideration of a de Sitter spacetime is justified in an epoch
where the energy density belonging to fluctuating degrees of freedom is sufficiently diluted as compared to the energy density of condensed degrees of freedom.

As shown in the Appendix, applying a Fierz transformation to the current-current interaction in (\ref{fermint}) yields
\begin{equation}\label{fierz}
   \left( \bar\psi\gamma_5\gamma_a\psi \right) 
   \left( \bar\psi\gamma_5\gamma^a\psi \right) 
   \to \frac{1}{N} \left( \bar\psi\psi \right)^2 
   + \frac{1}{N} \left( \bar\psi i\gamma_5\psi \right)^2 + \dots \,,
\end{equation}
where the dots refer to flavor-nonsinglet contributions and products of vector and axial-vector currents, which do not lead to vacuum condensates. Allowing for a bare cosmological constant $\Lambda_0$, and denoting the bare reduced Planck mass by $M_0$, the complete action then takes the form
\begin{equation}\label{Stot}
   S = \int d^4x\,e \left\{ M_0^2 H^2  - \Lambda_0
   + \bar\psi\,ie_a^\mu\gamma^a D_\mu[e]\,\psi + \frac{K}{2N} 
   \left[ \left( \bar\psi\psi \right)^2 
   + \left( \bar\psi i\gamma_5\psi \right)^2 + \dots \right] 
   \right\} ,
\end{equation}
where only the scalar flavor-singlet bilinears are shown explicitly. For condensation to take place $K$ needs to be positive, implying an imaginary $\gamma$ with $|\gamma|<1$. In fact, only if $K>0$ an attractive force occurs between fermions in the scalar channel \cite{njl}. In this case bound states form: The operator $\bar\psi\psi$ corresponds to a scalar field $\sigma$, while $\bar\psi i\gamma_5\psi$ corresponds to a pseudoscalar field $\pi$. In addition, $(N^2-1)$ flavor-nonsinglet scalar and pseudoscalar fields $\sigma^k\sim(\bar\psi t^k\psi)$ and $\pi^k\sim(\bar\psi t^k i\gamma_5\psi)$ appear; they are ignored in our discussion (see below). In the usual treatment of the Nambu--Jona-Lasinio model in Minkowski space \cite{njl} the dynamical breaking of chiral symmetry occurs for sufficiently large $K$. As a result, the scalar fields $\sigma$ and $\sigma^k $ become massive, while the pseudoscalar fields $\pi$ and $\pi^k$ remain massless and represent Goldstone bosons (see, e.g., \cite{Klevansky} for a review).

As we will see below, in a de Sitter background dynamical chiral symmetry breaking occurs for all values $K>0.$ Only the isosinglet fields are considered here. In fact, only the isosinglet scalar field $\sigma$ can acquire a nonzero vacuum expectation value and thus is relevant for de Sitter cosmology in the early Universe. Also, only the isosinglet pseudoscalar massless field $\pi$ can be associated with a Planck-scale axion field. It acquires a moderate mass by interacting with the topologically nontrivial field configurations of Yang-Mills theories in later epochs in the evolution of the Universe. In a de Sitter spacetime we expect that the particles associated with the remaining low-lying flavor-nonsinglet fields are sufficiently diluted to provide for the self-consistency of the de Sitter geometry. A detailed study of the validity of this assumption is beyond the scope of the present work.

On the technical side, the fields $\sigma$ and $\pi$ emerge when integrating out the fermions. They are first introduced as auxiliary fields in the action:
\begin{equation}\label{StotpB}
   S = \int d^4x\,e \left\{ M_0^2 H^2 - \Lambda_0 
   + \bar\psi\,ie_a^\mu\gamma^a D_\mu[e]\,\psi - \frac{N}{2K} 
   \left( \sigma^2 +\pi^2 \right) 
   - \bar\psi \left( \sigma + i\gamma_5 \pi \right) \psi \right\} .
\end{equation}
To lowest order in the $1/N$ expansion this is equivalent to the original action (\ref{Stot}) \cite{Inagaki1993}. We now introduce new variables $\rho$ and $\varphi$ as 
\begin{equation}\label{nV}
   \rho = \sqrt{\sigma^2+\pi^2} \,, \qquad 
   \tan\varphi =\frac{\pi}{\sigma} \,.
\end{equation}
A chiral rotation, which on $(\sigma,\pi)$ is represented by a fundamental $SO(2)$ rotation with angle $\alpha$, acts on the complex field $\chi\equiv\rho\,e^{i\varphi}$ in terms of a $U(1)$ phase shift: $\varphi\to\varphi+\alpha$. Upon integrating out the fermion fields $\psi$, the chirally symmetric effective potential $V_{\mathrm{eff}}$ depends on $\rho$ only, i.e., the Goldstone field $\varphi$ is precisely massless. The problem of computing the effective potential for a four-dimensional de Sitter spacetime was solved by Candelas and Raine using dimensional regularization \cite{CandelasRaine}.\footnote{Problems with defining $\gamma_5$ are absent due to the exact chiral symmetry.} 
An important observation made by these authors is that the ultra-violet divergences occurring in evaluating $\mbox{Tr}\,G(x,x;\rho)$, where $G(x,x;\rho)$ is the propagator of a fermion of mass $\rho$ in de Sitter spacetime, can be absorbed by renormalizing the bare parameters $\Lambda_0$ and $M_0$. This fact hints toward a deep link between four-dimensional gravity and the physics of gravitationally interacting, massless fermions in the large-$N$ limit. Recall that this limit is essential for the control of the mean-field (classical) treatment (\ref{StotpB}) of the action (\ref{Stot}). Note also that in Minkowski space no such connection exists: There a four-fermion interaction is nonrenormalizable, and one would have to introduce an ad hoc cutoff to tame divergent integrals.

We now discuss the renormalization of the parameters in the effective
Lagrangian in more detail, using dimensional regularization in $n=4+2\epsilon $ spacetime dimensions to regularize the ultra-violet 
divergences arising in the calculation. The starting point is the
dimensionally regularized action 
\begin{equation}\label{contact}
   S = \int d^{n}x\,e\left\{ 
   \frac{n(n-1)}{12}\,M_0^2 H^2 - \Lambda_0 
   + \bar\psi\,ie_{a}^{\mu} \gamma^{a} D_{\mu}[e]\,\psi 
   - \frac{N}{2K}\,\rho^{2}
   - \bar{\psi} \left( \sigma +i\gamma _{5}\pi \right) \psi 
   \right\} \,.  
\end{equation}
Following Candelas and Raine \cite{CandelasRaine}, who compute $\mbox{Tr}\,G(x,x;\rho)=-2e^{-1}\,\frac{\partial\mathcal{L}_{\mathrm{eff}}}{\partial\rho}$ associated with this action as 
\begin{equation}\label{tr}
   \mbox{Tr}\,G(x,x;\rho ) 
   = -2e^{-1}\,(2+\epsilon)\,
    \frac{\rho H^{2+2\epsilon}}{(4\pi)^{2+\epsilon }}\,
    \frac{\Gamma \left( 2+\epsilon +i\frac{\rho }{H}\right)
          \Gamma \left( 2+\epsilon -i\frac{\rho }{H}\right)}%
         {\Gamma \left( 1+i\frac{\rho }{H} \right) 
          \Gamma \left( 1-i\frac{\rho }{H}\right)}\,
     \Gamma(-1-\epsilon ) \,,
\end{equation}
the renormalized effective Lagrangian can be written in the form
\begin{equation}\label{Veffdef}
   e^{-1} \mathcal{L}_{\mathrm{eff}}
   = M_{\rm ren}^2(\mu)\,H^2 - \Lambda_{\mathrm{ren}}(\mu)
   - N\,V_{\mathrm{eff}}(\rho,H,\mu) \,,
\end{equation}
where the renormalized Planck mass $M_{\rm ren}$ and cosmological constant $\Lambda_{\rm ren}$ are defined so as to absorb the ultra-violet divergences arising in the calculation of the fermion determinant in the de Sitter background. This is accomplished by asymptotically expanding the expression for $e^{-1} \mathcal{L}_{\mathrm{eff}}$ in powers of $H^2$ and observing that divergent coefficients occur only at order $H^0$ and $H^2$. The resulting relations are\footnote{An additional divergence proportional to $H^4$ does not involve any fields and so has unobservable effects. Its subtraction is defined by setting the lower integration limit in (\ref{veff}) to zero.} 
\begin{eqnarray}\label{abs}
   \Lambda_0 &=& \mu^{2\epsilon} \left[
    \Lambda_{\mathrm{ren}}(\mu) + \frac{N\rho^4}{16\pi^2}
    \left( \frac{1}{\hat{\epsilon}} + c_{\Lambda} \right) 
    \right] , \notag \\
   M_0^2 &=& \mu^{2\epsilon} \left[
    M_{\rm ren}^2(\mu) + \frac{N\rho^2}{16\pi^2}
    \left( - \frac{2}{\hat{\epsilon}} + c_M \right) \right] ,  
\end{eqnarray}
where $1/\hat{\epsilon}\equiv 1/\epsilon+\gamma_E-\ln 4\pi$. The renormalization scale $\mu$ has been introduced so that the renormalized parameters have the canonical scaling dimensions $[\Lambda_{\mathrm{ren}}]=4$ and $[M_{\mathrm{ren}}^2]=2$. These parameters as well as the dimensionless quantities $c_\Lambda$, $c_M$ depend on the choice of the renormalization scheme. After eliminating the bare parameters in favor of the renormalized ones, the effective action remains finite in the limit $\epsilon\to 0$. The effective potential in (\ref{Veffdef}) is found to have the form
\begin{eqnarray}\label{veff}
   V_{\mathrm{eff}}(\rho,H,\mu ) 
   &=& \frac{\rho^{2}}{2K} + \frac{H^{4}}{8\pi^{2}}\,
    \Bigg\{\! \left( r^{2} + \frac{r^{4}}{2} \right) 
    \ln\frac{\mu^{2}}{H^{2}}
    - \left( \frac{1}{3} + \frac{c_{M}}{2}\right) r^{2}
    + \left( \frac{1}{4} + \frac{c_{\Lambda }}{2} \right) r^{4}
    \notag \\
   &&\mbox{}- 2\int_{0}^{r}\!dx\,x(1+x^{2})\,
    \Big[ \Psi(1+ix) + \Psi(1-ix) \Big] \Bigg\} \,,  
\end{eqnarray}
where $r=\rho/H$, and $\Psi(z)=d\log\Gamma(z)/dz$ denotes the digamma function. The effective potential depends on the Hubble parameter $H$ due to the presence of the de Sitter background. Note that the parameter $K$ (related to the Immirzi parameter) is not renormalized. 

As a brief interlude, it is instructive to work out the asymptotic behavior of the effective potential for small and large values of the ratio $r=\rho/H$. This will allow us to make contact with earlier discussions in the literature. For $r\ll 1$ (small $\rho $ at fixed $H$), we obtain 
\begin{eqnarray}\label{rklein}
   V_{\rm eff}(\rho,H,\mu) 
   &=& \frac{\rho^2}{2K}
    + \frac{1}{8\pi^2} \Bigg\{ 
    \rho^2 H^2 \left( \ln\frac{\mu^2}{H^2} + 2\gamma_E - \frac13 
    - \frac{c_M}{2} \right) \nonumber\\
   &&\mbox{}+ \rho^4 \left( \frac12\ln\frac{\mu^2}{H^2} + \gamma_E
    - \zeta_3 + \frac14 + \frac{c_\Lambda}{2} \right) 
    + \frac{2\rho^6}{3H^2}\,(\zeta_5-\zeta_3) + O(\rho^8/H^4) 
    \Bigg\} \,. \quad
\end{eqnarray}
The terms proportional to $H^2$ and $H^0$ (apart form the classical term $\frac{\rho^2}{2K}$) can be eliminated from the effective potential and absorbed into the renormalized parameters $M_{\mathrm{ren}}^2(\mu)$ and $\Lambda_{\mathrm{ren}}(\mu)$ if one sets $c_M=4\gamma_E-\frac23$ and $c_\Lambda=2(\zeta_3-\gamma_E)-\frac12$ and chooses the renormalization scale as $\mu=H$. This is the scheme choice adopted by Miao and Woodard \cite{Miao:2006pn}. Note that the renormalized parameters $\Lambda_{\mathrm{ren}}(H)$ and $G_{\mathrm{ren}}(H)$ now depend logarithmically on the Hubble parameter. In the opposite limit $r\gg 1$ (large $\rho$ for fixed $H$), we find 
\begin{eqnarray}
   V_{\rm eff}(\rho,H,\mu) 
   &=& \frac{\rho^2}{2K}
    + \frac{1}{8\pi^2} \Bigg\{ 
    \rho^4 \left( \frac12\ln\frac{\mu^2}{\rho^2} + \frac12 
    + \frac{c_\Lambda}{2} \right) 
    + \rho^2 H^2 \left( \ln\frac{\mu^2}{\rho^2} + \frac12 
    - \frac{c_M}{2} \right) \nonumber\\
   &&\mbox{}+ H^4 \left( - \frac{11}{60}\,\ln\frac{\rho^2}{H^2}
    + k \right) + \frac{31H^6}{1260\rho^2} + O(H^8/\rho^4) 
    \Bigg\} \,,
\end{eqnarray}
where $k\approx -0.6832$ is a (scheme-dependent \cite{CandelasRaine}) constant. Once again the terms proportional to $H^{0}$ (apart from the classical term) and $H^{2}$ can be eliminated from the effective potential and absorbed into the renormalized parameters. This is accomplished if one sets $c_\Lambda=-1$ and $c_M=1$ and chooses the renormalization scale as $\mu=\rho$. This scheme choice is adopted by Candelas and Raine \cite{CandelasRaine}. We stress that the physics is independent of these scheme and scale choices; they merely correspond to a reshuffling of terms between the effective potential and the renormalized parameters.

The definitions (\ref{abs}) imply the renormalization-group equations
\begin{equation}
   \frac{d\Lambda_{\rm ren}(\mu)}{d\ln\mu^2} 
    = - \frac{N\rho^4}{16\pi^2} \,, \qquad
   \frac{dM_{\rm ren}^2(\mu)}{d\ln\mu^2} = \frac{N\rho^2}{8\pi^2} \,,
\end{equation}
which show that the renormalized parameters $\Lambda_{\mathrm{ren}}(\mu)$ and $M_{\mathrm{ren}}^2(\mu)$ necessarily depend on $\rho$. In general this dependence cannot be calculated; to expose its precise nature would require an ultra-violet completion of the effective theory, in which divergences are regulated by dynamics rather than by an unphysical regulator. Dimensional analysis suggests that the $\rho$-dependent terms can be parameterized as
\begin{equation}\label{rhodep}
   \Lambda_{\rm ren}(\mu) = \Lambda - \frac{N\rho^4}{16\pi^2} 
    \left( \ln\frac{\mu^2}{\rho^2} + k_\Lambda \right) , \qquad
   M_{\rm ren}^2(\mu) = M_P^2 + \frac{N\rho^2}{8\pi^2}
    \left( \ln\frac{\mu^2}{\rho^2} + k_M \right) ,
\end{equation}
where $\Lambda$ and $M_P^2$ are $\rho$-independent. These relations account for the correct scale and scheme dependence if we require that the differences $c_i-k_i$ (for $i=\Lambda, M$) are scheme independent. The coefficients $k_i$ capture our ignorance about the ultra-violet completion of the model. The appearance of $\rho$ in the arguments of the logarithms is suggested by the fact that this is the only physical scale available. 

\subsection{Gap equation}

\begin{figure}
\begin{center}
\psfrag{x}[]{$\rho/M_P$}
\psfrag{y}[b]{$\bar V_{\rm eff}(\rho,H)/M_P^4$}
\includegraphics[width=0.5\textwidth]{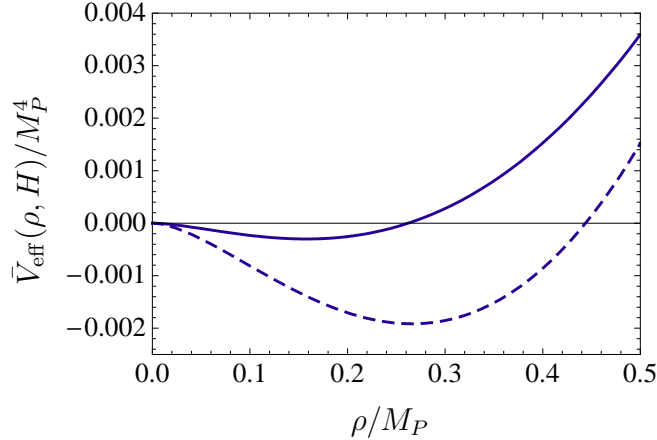} 
\end{center}
\vspace{-0.5cm}
\caption{
\label{fig:potential}
Renormalization-scheme invariant effective potential with parameter choices $H=M_P$, $K=20 M_P^{-2}$ (solid) and $H=1.5M_P$, $K=10 M_P^{-2}$ (dashed). We set $c_i=k_i=0$.}
\end{figure}

With our ansatz for the $\rho$-dependent terms in (\ref{rhodep}) we can now collect {\em all\/} $\rho$-dependent terms in the Lagrangian (\ref{Veffdef}) into the renormalization-scheme invariant effective potential
\begin{equation}\label{barVeff}
   \bar V_{\rm eff}(\rho,H)
   \equiv V_{\rm eff}(\rho,H,\mu) - \frac{\rho^4}{16\pi^2} 
   \left( \ln\frac{\mu^2}{\rho^2} + k_\Lambda \right) 
   - \frac{\rho^2 H^2}{8\pi^2}
   \left( \ln\frac{\mu^2}{\rho^2} + k_M \right) .
\end{equation}
The explicit expression for $\bar V_{\rm eff}(\rho,H)$ is obtained by performing the substitutions $\mu\to\rho$ and $c_i\to c_i-k_i$ in the expression for $V_{\rm eff}(\rho,H,\mu)$ in (\ref{veff}). The relevant asymptotic expansions read
\begin{equation}\label{smallr}
   \bar V_{\rm eff}(\rho,H) 
   = \frac{\rho^2}{2K} + \frac{\rho^2 H^2}{8\pi^2}  
    \left( \ln\frac{\rho^2}{H^2} + 2\gamma_E - \frac13 
    - \frac{c_M-k_M}{2} \right) + O(\rho^4)
\end{equation}
for $\rho\ll H$, and 
\begin{equation}
   \bar V_{\rm eff}(\rho,H) 
   = \frac{\rho^2}{2K} + \frac{\rho^4}{16\pi^2} 
   \left( 1 + c_\Lambda - k_\Lambda \right) + O(\rho^2 H^2) 
\end{equation}
for $\rho\gg H$. As long as $(c_\Lambda-k_\Lambda)>-1$ the potential is bounded from below and tends to infinity for $\rho\to\infty$. We must assume that this condition is satisfied in Nature. It can then be shown that the potential has a local maximum at $\rho=0$ and a single minimum at some positive value $\rho_{\rm min}$, which is determined by the solution of the equation
\begin{eqnarray}
   && \frac{4\pi^2}{K H^2} + \frac23 - \frac{c_M-k_M}{2}
    + \left( 1 + c_\Lambda - k_\Lambda \right) r_0^2
    \nonumber\\
   &=& (1+r_0^2) \left[ \Psi(1+ir_0) + \Psi(1-ir_0) - 2\ln r_0 
    \right] ; \qquad 
    r_0 = \frac{\rho_{\rm min}}{H} \,.
\end{eqnarray}
This minimum exists irrespective of the value and sign of $K$. This follows from the fact that the right-hand side is a monotonically decreasing function of $r_0$ starting at infinity for $r_0=0$, while the left-hand side is monotonically increasing and tends to infinity for $r_0\to\infty$. Figure~\ref{fig:potential} shows the shape of the effective potential for two sets of parameters.

\begin{figure}
\begin{center}
\psfrag{x}[]{$KH^2$}
\psfrag{y}[]{$\rho_{\rm min}\sqrt{|K|}$}
\includegraphics[width=0.5\textwidth]{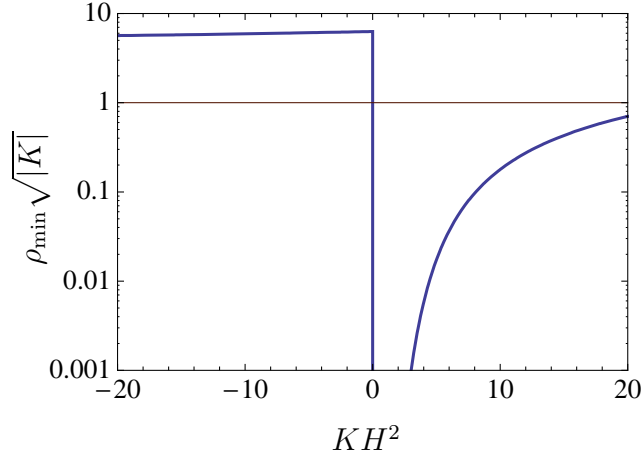} 
\end{center}
\vspace{-0.5cm}
\caption{
\label{fig:minimum}
Position of the minimum of the renormalization-scheme invariant effective potential as a function of $K H^2$. We set $c_i=k_i=0$. The branch with $K H^2<0$ is unphysical.}
\end{figure}

At this point an important remark is in order. For consistency of the de Sitter calculation \cite{CandelasRaine} the scalar field $\rho$ needs to be homogeneous, such that the effective Lagrangian does not contain a kinetic term for $\rho$. Note that while expression (\ref{StotpB}) does not contain such a term, it can in principle be generated by the leading term in the derivative expansion of the fermion determinant about an arbitrary field configuration $\rho(x)$. Here we will assume that after the fermions condense the dynamics   quickly places the scalar field at the minimum of the potential. With this assumption our derivation is self-consistent.

The gap equation follows from the stationary condition for the effective potential (\ref{barVeff}) under variations of $\rho$: 
\begin{equation}
   \frac{\partial\mathcal{L}_{\mathrm{eff}}}{\partial\rho}
   \overset{!}{=} 0 \quad \Rightarrow \quad
   \frac{\partial\bar V_{\mathrm{eff}}(\rho,H)}{\partial\rho} = 0 \,.
\end{equation}
Figure~\ref{fig:minimum} shows the position of the minimum of the effective potential as a function of the product $K H^2$. For negative $K$ the minimum is located at large $\rho_{\rm min}=O(2\pi/\sqrt{-K})$ and is approximately independent of $H$. This branch is unphysical. It leads to a large (Planck-scale) negative potential energy, which for $N\to\infty$ is incompatible with a de Sitter Universe (see the Friedmann equation~(\ref{fried}) below). For positive $K$, on the other hand, the minimum is located at small values of $\rho/H$. Keeping the first two terms in the expansion (\ref{smallr}), we obtain
\begin{equation}\label{apprhoklH}
   \rho_{\rm min}^2 
   = H^2 \exp\left( - \frac{4\pi^2}{K H^2} - C \right) ,    
\end{equation}
where we have defined the scheme-independent constant $C=2\gamma_E+\frac23+\frac{k_M-c_M}{2}$. Thus a vacuum expectation value $\rho _{\mathrm{min}}\neq 0$ develops for each positive value of $K$, contrary to the Minkowski case, where condensation occurs only if $K$ exceeds a critical value \cite{Klevansky}. Under the reasonable assumption that $K H^2\ll 4\pi^2$ we find that $\rho_{\mathrm{min}}\ll H$. This is consistent with the determination of the minimum of the effective potential using the approximated form (\ref{smallr}). The minimum value is found to be
\begin{equation}\label{Vmin}
   \bar V_{\mathrm{eff}}(\rho_{\rm min},H)
   = - \frac{\rho_{\rm min}^2 H^2}{8\pi^2} + O(\rho_{\rm min}^4) \,.
\end{equation}

\section{Cosmological implications}
\label{cosmo}

\subsection{Friedmann equation and relaxation of the expansion rate}
\label{relL}

The Friedmann equation for de Sitter spacetime subject to the renormalized cosmological constant $\Lambda$ and the energy density of the condensed scalar field $\rho$, resulting from the process of integrating out the interacting fermions, reads 
\begin{equation}\label{fried}
   H^2 = \frac{1}{M_P^2} 
   \left( \Lambda + N \bar V_{\mathrm{eff}}(\rho_{\rm min},H) 
   \right) .
\end{equation}
Introducing the dimensionless variables
\begin{equation}
   h = \frac{H}{M_P} \,, \qquad 
   \lambda = \frac{\Lambda}{M_P^4} \,, \qquad 
   k = K M_P^2 \,, \qquad
   \bar N = N\,e^{-C} \,,
\end{equation}
we obtain with (\ref{apprhoklH}) and (\ref{Vmin})
\begin{equation}\label{final}
   \lambda= h^2 + \frac{\bar N h^4}{8\pi^2}\,
   \exp\left( - \frac{4\pi^2}{k h^2} \right) .
\end{equation}
Note that our ignorance about the ultra-violet completion of the model hides in a harmless $O(1)$ rescaling of $N$. It is natural to assume that in these Planck units $\lambda,k=O(1)$. Relation~(\ref{final}) would then predict $h\approx\lambda$ for $N=O(1)$. On the other hand, in the limit $N\to\infty$, which is required for the self-consistency of our approach, the same relation implies that $h\to 0$, independent of the precise values of $\lambda $ and $k$. In other words, a relaxation of Hubble expansion to zero takes place for very large $N$. 

The question of what happens if we add a matter sector, which introduces a substantial deviation from de Sitter cosmology, is open for two reasons: First, the renormalizability of our four-fermion model may rely crucially on the large symmetry of de Sitter spacetime \cite{CandelasRaine}, and it is not clear which interacting field theories keep their predictivity on more general geometries. On a Minkowski spacetime we know that these theories are of the Yang-Mills type, and we would expect that their predictivity does not get spoiled by mild deformations of this particular background. Second, we have no easy analytical handle on a general matter sector (subject to a product of finite, nonabelian gauge groups) in conjunction with the above fermion model, although we believe that future investigations will gain deeper insights into this issue. An interesting scenario would be that the relaxation of Planckian vacuum energy density to zero is a step which precedes the liberation of finitely many gauge symmetries being responsible for subsequent cosmology.

\subsection{A Planck-scale axion}
\label{PSA}

Let us now turn to the isosinglet Goldstone field $\varphi$, which plays the role of a Planck-scale axion \cite{frieman1995} if, by means of the chiral anomaly \cite{Adler,JackiwBell}, it couples to Yang-Mills theories in phases with propagating gauge modes \cite{GH2005}. The interaction of the canonically normalized field $\varphi$ with a Yang-Mills theory is described by\footnote{For various reasons, which we may elaborate on in the future, it appears natural that chiral fermions and gravitation emerge at the Planck scale from a confining $SU(N=\infty)$ Yang-Mills theory. The Planck mass $M_P$ would then naturally act as the Peccei-Quinn scale.} 
\begin{equation}\label{couplaxion}
   \mathcal{L}_{\mathrm{int}}
   \sim \frac{\varphi}{M_P}\,\mbox{tr}\,\tilde{F}_{\mu\nu } 
   F^{\mu\nu} \,.  
\end{equation}
Recall that the chiral anomaly occurs on top of a dynamical breakdown of the global $U_A(N)$ symmetry carried by the fermions $\psi$. Integrating out (in addition to the fermions) the ground-state portion of the gauge field $F_{\mu\nu}^i$, the field $\varphi$ acquires a potential of the form \cite{Wilczek,Peccei} 
\begin{equation}
   \left( 1 - \cos\frac{\varphi}{M_P} \right) \Lambda_{\rm YM}^4 \,,
\end{equation}
where $\Lambda_{\rm YM}$ refers to the Yang-Mills scale. The field $\varphi$, which naturally emerges in the context of our model, can be interpreted as a Planck-scale axion responsible for the late-time evolution of the Universe \cite{GH2005,frieman1995}. Thus, within our approach, the physics of the relaxation of Planck-scale vacuum energy is ultimately connected with the physics of dark energy today. This relates point~A with point~C in the Introduction.

It was pointed out in \cite{Enqvist:2001zp} that an oscillating axion field in FRW cosmology eventually will dominate the energy density of a formerly radiation-dominated Universe, and that, upon decay into photons, its fluctuations induce CMB anisotropies that no longer are of a purely isocurvature nature.

\section{Conclusions}
\label{SC}

We have developed a scenario for the relaxation of Planckian vacuum energy under the assumption that $N$ chiral fermions (microscopically) interact via the torsion term (\ref{induced}) arising from the Holst action of general relativity. In the large-$N$ limit, for which the model is renormalizable in a de Sitter geometry \cite{CandelasRaine}, the Hubble constant $H$ vanishes for any fixed value of the renormalized cosmological constant. After integrating out the fermionic degrees of freedom we find that two composite isosinglet fields play a crucial role: a scalar field $\rho$ is relevant for the early-time cosmology and for the above-mentioned relaxation of the vacuum energy, while a pseudoscalar field $\varphi$ emerges as an axion field, which can play a crucial role for late-time cosmology, being responsible for the presently observed small but nonzero value of the dark energy \cite{Hofmann2005,GH2005,frieman1995,Hofmann:2005qk}.

In the evaluation of the effective potential for early cosmology care is needed in the choice of the regularization scheme. Lacking a definite ultra-violet completion, a dependence of the effective potential on arbitrary renormalization constants persists. In addition, the renormalized Planck mass and cosmological constant depend on the condensate $\rho$. We have elucidated how previous discussions in the literature \cite{CandelasRaine,Miao:2006pn} correspond to different choices of renormalization schemes. Fortunately, the small-$\rho$ behavior (\ref{smallr}) of the effective potential, which is important for our relaxation mechanism, is not significantly affected by these considerations. Our ignorance about the values of the renormalization constants can be absorbed into an $O(1)$ rescaling of $N$.

It is important to stress that the minimum of the effective potential $\bar V_{\mathrm{eff}}(\rho,H)$ occurs for $\rho_{\mathrm{min}}\neq 0$ as long as the parameter $K$, which measures the strength of the four-fermion interaction, is positive. Thus condensation always takes place, contrary to the case of Minkowski spacetime, for which the interaction strength must exceed a critical value. This implies the existence of the (axion) field $\varphi$ for all times. Since for a sufficiently small number of independent Yang-Mills scales ($\Lambda_{\rm YM}\ll M_P$) for the gauge dynamics governing the nongravitational sector the mass of the field $\varphi$ is much smaller than $M_P$ (the scale of $H$ during inflation), this field may in addition serve as a curvaton \cite{Lyth:2001nq}. That is, an almost scale-invariant spectrum of isocurvature perturbations, imprinted into $\varphi$ on super-horizon scales in the final stage of inflation, triggers curvature perturbations and thus large-scale structure formation upon its re-entry into the horizon during a much later epoch. That the role of the curvaton can be played by a (pseudo) Goldstone boson of a dynamical chiral symmetry breaking was discussed previously in \cite{Enqvist:2001zp,Dimopoulos:2003ii,Hofmann:2002gy}. We leave an analysis of this and other interesting problems, such as the study of tunneling into anti-de Sitter spacetime (in the case where the potential is unbounded from below) and the cosmological evolution after inflation, for future study.

\subsection*{Acknowledgments}

Two of us (FG and RH) would like to acknowledge very useful discussions with Tomislav Prokopec at an early stage of this work.

\section*{Appendix: Fierz transformation}

The gravitationally-induced four-fermion interaction can be rewritten in the form
\begin{eqnarray}
   \sum_{i,j=1}^N
    \left( \bar\psi_i\gamma_5\gamma^\mu\psi_i \right)
    \left( \bar\psi_j\gamma_5\gamma_\mu\psi_j \right)
   &=& \sum_{i,j=1}^N \bigg[
    \left( \bar\psi_i\psi_j \right) \left( \bar\psi_j\psi_i \right)
    + \left( \bar\psi_i i\gamma_5\psi_j \right)
    \left( \bar\psi_j i\gamma_5\psi_i \right) \nonumber\\
   &&\mbox{}+ \frac12 \left( \bar\psi_i\gamma^\mu\psi_j \right)
    \left( \bar\psi_j\gamma_\mu\psi_i \right)
    + \frac12 \left( \bar\psi_i\gamma_5\gamma^\mu\psi_j \right)
     \left( \bar\psi_j\gamma_5\gamma_\mu\psi_i \right)
   \bigg] \,, \qquad \nonumber
\end{eqnarray}
where $i,j$ are flavor indices. We now introduce the traceless generators $t_A$ of $SU(N)$ (with $A=1,\dots,N^2-1$), normalized such that $\mbox{Tr}(t_A t_B)=\frac12\delta_{AB}$. They obey the relation
\[
   \sum_A\,(t_A)_{ij} (t_A)_{kl} = \frac12\,\delta_{il}\delta_{kj}
   - \frac{1}{2N}\,\delta_{ij}\delta_{kl} \,.
\]
This can be used to rewrite the flavor structure of the operators on the right-hand side of the above relation in the form
\[
   \left( \bar\psi_i\psi_j \right) \left( \bar\psi_j\psi_i \right)
   = \frac{1}{N} \left( \bar\psi\psi \right)
    \left( \bar\psi\psi \right)
   + \sum_A \left( \bar\psi\,t_A \psi \right)
    \left( \bar\psi\,t_A\psi \right) ,
\]
and similarly for the other Lorentz structures.

\end{document}